\documentclass[showpacs,twocolumn,prl]{revtex4}
\usepackage{amsmath}

\DeclareMathOperator{\prob}{Prob}

\begin{document}

\title{An exactly soluble noisy traveling wave equation appearing in the
problem of directed polymers in a random medium}
\author{\'Eric Brunet}\email{Eric.Brunet@lps.ens.fr}
\author{Bernard Derrida}\email{Bernard.Derrida@lps.ens.fr}
\affiliation{Laboratoire de Physique Statistique, \'Ecole Normale
Sup\'erieure, 24 rue Lhomond, 75231
	Paris Cedex 05, France}

\begin{abstract}
We calculate exactly the velocity and diffusion constant of a
microscopic stochastic model of $N$ evolving particles which can be
described by a noisy traveling wave equation with a noise of order
$N^{-1/2}$. Our model can be viewed as the infinite range limit of a
directed polymer in random medium with $N$~sites in the transverse
direction. Despite some peculiarities of the traveling wave equations in
the absence of noise, our exact solution allows us to test the validity
of a simple cutoff approximation and to show that, in the weak noise
limit, the position of the front can be completely described by the
effect of the noise on the first particle. 
\end{abstract}

\maketitle

Traveling wave equations such as the
Fisher-Kol\-mo\-go\-rov-Petrovsky-Piscounov (F-KPP) equation
\cite{Fisher.37, KPP.37, Bramson.83}
\begin{equation}
\partial_t u = \partial_x^2 u + u - u^2
\label{FKPP}
\end{equation}
describe moving fronts \cite{vanSaarloos.03} in a large number of
problems in biology, chemistry and genetics. In physics, it
appears in non-equilibrium statistical mechanics and in the theory of
disordered systems \cite{DerridaSpohn.88, CarpentierLeDoussal.00}. In
typical cases, $u_t(x)$ represents the concentration at position $x$ and
time $t$ of some chemical species or of individuals carrying a certain
gene. It is well established that equations of the F-KPP type have a one
parameter family of traveling wave solutions with $u(-\infty)=1$ and
$u(\infty)=0$ parametrized by their velocities~$v$, and that the decay of
the initial condition determines the velocity of the front. For localized
initial conditions, the velocity is the minimal velocity allowed
$v_\text{min}=2$ \cite{AronsonWeinberger.75, Bramson.83, vanSaarloos.03}.

When deriving traveling wave equations such as (\ref{FKPP}) from a given
microscopic stochastic models \cite{Bramson.86, Kerstein.86,
BrunetDerrida.97, MuellerSowers.95, PechenikLevine.99, Moro.03}, one usually gets
a noisy version of (\ref{FKPP}):
\begin{equation}
\partial_t u = \partial_x^2 u + u - u^2 + \sqrt{\epsilon}\, g(u)\, \xi(x,t),
\label{FKPP1}
\end{equation}
where the additional term is proportional to a function $g(u)$ of the
concentration and to a Gaussian noise $\xi(x,t)$ white in time and
possibly correlated in space. The amplitude $\epsilon$ represents the
ratio between the microscopic and the macroscopic scales. Internal
fluctuations due to the finite number of interacting particles gives
\cite{MuellerSowers.95, Karzazi.96, PechenikLevine.99,
DoeringMuellerSmereka.03} $g(u)\propto\sqrt {u(1-u)}$ . Although it is
now established \cite{MuellerSowers.95} that the presence of the noise
term is sufficient \cite{Bramson.86, Kerstein.86, Mai.96,
BrunetDerrida.97} to select a single velocity $v_\epsilon$, and that this
selected velocity $v_\epsilon$ tends slowly \cite{Kerstein.86,
BrunetDerrida.97} in the limit $\epsilon \to 0$ to the minimal velocity
$v_{\rm min} $ allowed by (\ref{FKPP}), it is still a theoretical challenge
\cite{BrunetDerrida.97, Kessler.98, PechenikLevine.99, vanSaarloos.03,
Panja.03} to predict how $v_{\rm min} - v_\epsilon$ vanishes with
$\epsilon$ or how specific properties of the noisy equation such as the
diffusion constant $D_\epsilon$ of the front position behave in
the small noise limit. Other types of noise have also been considered;
for instance, taking $g(u)\propto u$ in (\ref{FKPP1}) would represent the
fluctuations of a control parameter due to some external
noise\cite{Armero.98, RoccoEbertvanSaarloos.00}. 

Here, we consider a microscopic model which can be viewed as the problem
of directed polymers in a random medium \cite{Halpin-HealyZhang.95} where
we take the infinite range limit in the transverse direction. As
explained below, this problem leads to a noisy traveling wave equation
and, for one specific choice of the disorder in the directed polymer
problem, we can solve the microscopic dynamics and calculate exactly the
velocity, diffusion constant and all the higher cumulants of the position
of the front. This exact solution allows us to test several approximation
schemes used recently to attack the general problem of noisy fronts.

The microscopic model we consider here is defined as follows: at each time
step~$t$ we have $N$ particles on the real axis at positions
$x_1(t),\dots, x_N(t)$. Given these positions at time~$t$, the new
positions at time~$t+1$ are obtained by
\begin{equation}
 x_i(t+1) = \max_{1 \leq j \leq N} [ x_j(t) + s_{i,j}(t) ]
\label{def-model}
\end{equation}
where the $s_{i,j}(t)$ are independent random numbers generated
according to a given distribution~$\rho(s)$. One can think of $-x_i(t)$
as being the ground state energy of a directed polymer of length~$t$
ending at position~$i$ in the ``transverse'' direction.
Then~(\ref{def-model}) describes the infinite range case in the transverse
direction as the directed polymer may jump at any step~$t$ from any
site~$j$ to any site~$i$ and gain a bond energy $-s_{i,j}(t)$. It is
easy to show that the points remain grouped and we want to know the
velocity $v_\text{exact}$ and the diffusion $D_\text{exact}$ of this
cloud of points (or, equivalently, of its center of mass).

One can associate a traveling wave equation to (\ref{def-model}) by
considering the proportion~$u_t(x)$ of particles on the right of~$x$:
\begin{equation}
u_t(x) = \frac{1}{N} \sum_{1\le i\le N} \theta\big(x_i(t)-x\big),
\label{utdef}
\end{equation}
where $\theta(z)$ is the Heaviside function. Clearly, $u_t(x)$ is a
decreasing function with $u_t(-\infty)=1$ and $u_t(\infty)=0$, so that
$u_t(x)$ has the shape of a front.

For a given front configuration~$u_t(x)$, the random
variables~$\{x_i(t+1)\}$ are uncorrelated and the probability to
be on the right of~$x$ is given by
\begin{equation}
\prob\big(x_i(t+1)>x\big)=1-\prod_{1\le j\le N} \mu(x-x_j(t)),
\label{pdef}
\end{equation}
where $\mu(s)$ is the probability that~$s_{ij}(t)<s$:
\begin{equation}
\mu(s) = \int_{-\infty}^s \rho(s')\ ds'		 \label{mu-def}
\end{equation}

For a given~$u_t(x)$, the probability (\ref{pdef}) is simply the
front~$\langle u_{t+1}(x)\rangle$ at time~$t+1$ averaged \emph{over one
time step}. One could rewrite (\ref{pdef}) using only the front
variable~$u_t(x)$:
\begin{equation}
\big\langle u_{t+1}(x)\big\rangle=1-\exp\left[{-N\int
u_t(y)\frac{\mu'(x-y)}{\mu(x-y)}\ dy}\right],
\label{noisy1}
\end{equation}
but one needs to be careful when~$\mu$ vanishes and use the
following prescription: When~$\mu(x-y)=0$, then $u_t(y)\mu'(x-y)/\mu(x-y)$
should be set to 0 if~$u_t(y)=0$ and to $+\infty$ if~$u_t(y)\ne0$.

The random positions~$x_i(t+1)$ can be generated by
solving~$\big\langle u_{t+1}(x_i)\big\rangle=z_i$, where $z_1, \dots,
z_N$ are $N$ independent random numbers uniformly chosen between 0
and 1 and the fluctuating front~$u_{t+1}(x)$ is finally given by
\begin{equation}
u_{t+1}(x)= \frac{1}{N}\sum_{1\le i\le N} 
	\theta\Big( \big\langle u_{t+1}(x) \big\rangle - z_i \Big).
\label{noisy2}
\end{equation}
From (\ref{noisy2}), one can calculate how the fluctuations of
$u_{t+1}(x)$ are correlated. Writing
\begin{equation}
u_{t+1} (x) = \big\langle u_{t+1} (x) \big\rangle + 
\frac{1}{\sqrt{N}}\ \eta_{t+1}(x),
\label{ut-noisy}
\end{equation}
one finds for $x \leq y$
\begin{equation}
\big\langle \eta_{t+1}(x) \eta_{t+1}(y) \big\rangle = 
 \Big(1- \big\langle u_{t+1}(x)\big\rangle \Big)
\big\langle u_{t+1}(y) \big\rangle
\label{c2}
\end{equation}
Writing~$\eta(x)=\xi(x)\sqrt{\langle\eta^2(x)\rangle}$, Eq.~(\ref{ut-noisy})
becomes very similar to (\ref{FKPP1}) with~$g(u)=\sqrt{u(1-u)}$. So far,
(\ref{noisy1}, \ref{noisy2}) and their consequence (\ref{ut-noisy},
\ref{c2}) are exact, for arbitrary $N$. From (\ref{noisy2}) one can also
calculate higher correlations of $\eta_{t+1}(x)$ and show that,
for large~$N$, the $\eta$ become Gaussian. For example one can
check that, up to terms of order~$1/N$,
\begin{displaymath}
\langle \eta_1 \eta_2 \eta_3 \eta_4 \rangle 
= \langle \eta_1 \eta_2 \rangle \langle \eta_3 \eta_4 \rangle 
 +\langle \eta_1 \eta_3 \rangle \langle \eta_2 \eta_4 \rangle 
 +\langle \eta_1 \eta_4 \rangle \langle \eta_3 \eta_4 \rangle 
\end{displaymath}
(We used the simplified notation $\eta_i \equiv \eta_{t+1}(x_i)$.)
We should notice that the Gaussian character of the $\eta_{t+1}(x)$ is
a property valid only for large $N$ and in regions where
\begin{equation}
N \Big(1- \big\langle u_{t+1}(x)\big\rangle \Big) \gg 1 \quad 
\text{and} \quad 
N \big\langle u_{t+1}(x)\big\rangle \gg 1.
\label{validity}
\end{equation}

From now on, we will limit our discussion to the case where $\rho(s)$
is a Gumbel distribution:
\begin{equation}
	\rho(s)= \exp\left(-s -e^{- s} \right).\label{gumbel}
\end{equation}
In that case, the full analysis of the problem becomes easy and we can
calculate exactly the statistical properties of the front in the large
$N$ limit. From (\ref{mu-def}) and \r{gumbel},
one has $\mu'(s) / \mu(s) = e^{-s}$ and (\ref{noisy1}) becomes 
\begin{equation}
\label{noisy1-gumbel}
\langle u_{t+1} (x) \rangle= 1 - \exp\left[- B_t \ e^{- x} \right],
\end{equation}
where $B_t$ is defined by
\begin{equation}
\label{Bt-def}
B_t = N \int e^x \, u_t(x) \ dx = \sum_{1\le i \le N} e^{x_i(t)}.
\end{equation}
This definition and (\ref{noisy1-gumbel}) imply that, given $B_t$,
 \begin{equation}
\langle B_{t+1} \rangle =\infty,
\label{Btp1infinite}
\end{equation}
which means that the distribution of the random variable $B_{t+1}$ decays
slowly.

The main advantage of the Gumbel distribution~(\ref{gumbel}) is that the
maximum of several Gumbel variables is itself distributed according
to a Gumbel distribution. Therefore, for fixed $x_i(t)$ in~(\ref{def-model}),
having the~$s_{i,j}(t)$ distributed according to the Gumbel distribution
(\ref{gumbel}) implies that~$x_{i+1}(t)$ is itself a Gumbel variable.

The Gumbel distribution is simple because $B_t$ in~(\ref{Bt-def}) is the only
information needed to construct the front $u_{t+1}(x)$ at time $t+1$. If
one defines the position $X_t$ of the front at time $t$ by $X_t = \ln
B_t$, then the displacements
\begin{equation}
\Delta X_t = X_{t+1} - X_{t} = \ln B_{t+1}-\ln B_t
\label{DeltaXt-def}
\end{equation}
are uncorrelated random variables given by
\begin{equation}
\Delta X_t = \ln \left[ \frac{1}{y_{1}(t+1) } + \cdots +
\frac{1}{ y_{N}(t+1) } \right],
\label{Yt-expression}
\end{equation}
where the $y_i(t+1)=B_t \exp[- x_i(t+1)]$ are independent and, using
(\ref{pdef}, \ref{noisy1}, \ref{noisy1-gumbel}), distributed
according to 
\begin{equation}
p(y)= e^{-y} \ \theta(y).
\label{rho(y)}
\end{equation}
As the $\Delta X_t$ are independent, the cumulants of the position $X_t$
are simply~$t$ times those of~$\Delta X_t$, which can be calculated
from its generating function. The identity
\begin{equation}
\big\langle e^{- \delta \Delta X_t} \big\rangle =
\frac{1}{\Gamma(\delta)}\int_0^\infty du\ u^{\delta -1}
\left\langle \exp \left( - u e^{\Delta X_t} \right) \right\rangle,
\label{integral-representation}
\end{equation}
and the fact that the $y_i(t)$ are independently distributed according
to (\ref{rho(y)}), lead to the following exact expression:
\begin{equation}
\big\langle e^{- \delta \Delta X_t} \big\rangle
 = \frac{1}{\Gamma(\delta)}\int_0^\infty du\ u^{\delta -1} 
\left( \int_0^\infty e^{-y - \frac{u}{y}}\ dy \right)^N.
\label{exact-generating-function}
\end{equation}
For large $N$, the integral over $u$ is dominated by the neighborhood of $u=0$
where
\begin{equation}
 \int_0^\infty e^{-y - \frac{u}{y}}\ dy
 \simeq 1 + u \ln u + (2\gamma_E-1)u + O\big(u^2 \ln u \big) ,
\label{asymptotics}
\end{equation}
where~$\gamma_E=-\Gamma'(1)=-\int_0^{+\infty}du\ e^{-u}\ln u$ is the
Euler gamma constant. Any term of higher order in~$u$ would give a
correction $1/N$ to the final result. Using (\ref{asymptotics}) into
(\ref{exact-generating-function}), one obtains, for large $N$,
\begin{multline}
\ln \langle e^{- \delta \Delta X_t} \rangle = 
-\delta \left(
L + \ln L \right)\\
-\frac{\delta}{L}\left[\ln L + 1 - 2 \gamma_E
 - \frac{\Gamma'(1+ \delta)} {\Gamma(1 + \delta)}
\right] + o \left( \frac{1}{L} \right),
\label{result}
\end{multline}
where $L \equiv \ln N $.
This leads to the following exact expressions for the velocity and of the diffusion constant of the front described by
(\ref{noisy1}--\ref{c2}) when $\rho(s)$ is given by (\ref{gumbel}):
\begin{align}
&v_\text{exact}=\lim_{t\to \infty}\frac{\langle X_t \rangle}{t} = L +
\ln L +\frac{ \ln L}{L } + \frac{1 - \gamma_E}{ L} + o\left(\frac{1}{L}
\right) \notag\\ 
&D_{\rm exact}=\lim_{t \to \infty}\frac{\langle X_t^2 \rangle - \langle
X_t \rangle^2}{t}= \frac{\pi^2}{3 L} + o\left(\frac{1}{L} \right).
\label{exact}\end{align}
Expanding~(\ref{result}) in powers of~$\delta$ gives also all the higher
cumulants of the position of the front.

Armed with these exact results, one can test the quality of various
approximation schemes\cite{BrunetDerrida.97,BrunetDerrida.01}.

A first approximation is to neglect the noise and to write
a deterministic traveling wave equation for the front. However, if one
replaces~$\langle u_{t+1}(x)\rangle$ by $u_{t+1}(x)$ in (\ref{noisy1}), one
obtains for~(\ref{gumbel}) a meaningless front equation: starting with
$u_0(x)=\theta(-x)$, one finds~$u_2(x)=1$ for all~$x$. Equation
(\ref{noisy1}) is meaningful only in the presence of noise.

Another way of
removing the noise is to assume that for each position~$x_i(t+1)$
in~(\ref{pdef}), all the $x_j(t)$ are uncorrelated random variables chosen
independently for each~$i$. This leads to the following deterministic
equation of a front propagating into an unstable state:
\begin{equation}
u_{t+1}(x)=1-\left[1-\int dy\ u_t(y)\rho(x-y)\right]^N.
\label{thermo}
\end{equation}
This equation is very similar to~(\ref{FKPP}) and its velocity can be
obtained using the usual method\cite{vanSaarloos.03}: looking for
solutions of the form~$u_t(x)=\exp[-\gamma (x-v(\gamma) t)]$
when~$u_t(x)\ll1$, one
obtains a function $v(\gamma)$. For initial conditions which decay fast
enough, the velocity is $v_\text{meanfield}=\min_\gamma v(\gamma)$.
When~$\rho(s)$ is the Gumbel distribution, one gets $v(\gamma)=[\ln
N+\ln\Gamma(1-\gamma)]/\gamma$ and, for large~$N$,
the minimal velocity is $v_\text{meanfield}=v_\text{exact}+1+o(1/L)$. One
could try to improve this result by using the cutoff
approximation\cite{BrunetDerrida.97,
Kessler.98, BrunetDerrida.01} in (\ref{thermo}), but as~$v(\gamma)$
depends on~$N$, it is not clear that the formula $\Delta
v=-\pi^2\gamma^2v''(\gamma)/(2\ln^2N)$ derived in \cite{BrunetDerrida.97}
can be applied. Trying to apply it anyway,
one obtains the velocity~$v_\text{meanfield}-\pi^2/2$ which is not closer
to~$v_\text{exact}$ than $v_\text{meanfield}$.

The cutoff approximation can also be applied directly on
the evolution equations~(\ref{noisy1-gumbel}, \ref{Bt-def})
by setting $u_{t+1}(x)=\langle u_{t+1}(x)\rangle$ whenever $\langle
u_{t+1}(x)\rangle>\lambda/N$ and $u_{t+1}(x)=0$ otherwise.
One can write a closed expression for the evolution
of~$B_t$:
\begin{equation}
B_{t +1}= N \int_{-\infty}^{A_{t+1}} e^{ x} \left(1 - e^{-B_t
e^{- x}} \right)\ dx ,
\label{Bt-cutoff}
\end{equation}
where the position $A_{t+1}$ of the cutoff is given by
\begin{equation}
\frac{\lambda}{N}= \left\langle u_{t+1}(A_{t+1})\right\rangle=1 - e^{-B_t e^{- A_{t+1}}}.
\label{cutoff-def}
\end{equation}
By eliminating $A_{t+1}$, one gets
\begin{equation}
B_{t+1}= B_t \ N \int_{\ln \frac{N}{N-\lambda}}^\infty\frac{1-
e^{-u}}{ u^2}\ du.
\label{Bt-evolution}
\end{equation}
Using $v_\text{cutoff}=\ln (B_{t+1}/ B_t)$, we get, for
large~$L\equiv\ln N$
\begin{equation}
v_{\rm cutoff}= L + \ln L + \frac{1 - \gamma_E -\ln\lambda}{L }
+o\left(\frac{1}{L}\right).
\label{vcutof1}
\end{equation}
Comparing (\ref{vcutof1}) and \r{exact}, we see that $v_{\rm cutoff}$ gives
correctly the leading orders in $L$  with a discrepancy $(\ln L)/L$,
which is only slightly above what the cutoff
approximation may predict anyway, as there is no reason to choose any
particular value of $\lambda$, and which is much better than the
mean-field velocity obtained from~(\ref{thermo}).

So far, all the approximations replaced the noisy dynamics by a
deterministic equation. Thus, they gave no prediction for the
diffusion constant. We will now examine approximations in which the
system remains noisy.

A first possibility is to consider that the evolution is given by
(\ref{ut-noisy}, \ref{c2}, \ref{noisy1-gumbel}, \ref{Bt-def}) where
the noise term in (\ref{ut-noisy}) is exactly Gaussian, even outside the
validity range~(\ref{validity}). In this case, $B_{t+1}$ is Gaussian but
with $\langle B_{t+1} \rangle = \infty$ (see (\ref{Btp1infinite})). The noisy
front equation (\ref{ut-noisy}) becomes meaningless after a single time
step. So, one cannot ignore that the noise is not Gaussian near the
rightmost particle.

The next approximation one can try is to put a cutoff into the system as
in (\ref{cutoff-def}), and to take into account the effect of all
the fluctuations on the left of this cutoff. In other words, at each time
step, we determine $A_{t+1}$ by (\ref{cutoff-def}), we set $u_{t+1}(x)=0$ for 
$x > A_{t+1}$ and we use (\ref{ut-noisy}) for $x < A_{t+1}$ with a
\emph{Gaussian} noise $\eta_{t+1}(x)$ correlated as in (\ref{c2}). This leads
to a Gaussian $B_{t+1}$ characterized by $\langle B_{t+1} \rangle $
given by the right hand side of (\ref{Bt-cutoff}) or \r{Bt-evolution} and
\begin{multline}
\langle B_{t+1}^2 \rangle - \langle B_{t+1} \rangle^2 = 2 N
\int_{-\infty}^{A_{t+1}}dy\ ( 1 - e^{-B_t e^{- y}}) e^{ y}\\
\times \int_{-\infty}^y dx\ e^{ x} e^{-B_t e^{- x}}.
\label{approx2}
\end{multline}
For large $N$, one easily gets
$\langle B_{t+1} \rangle \simeq N B_t \ln N$ and
$\langle B_{t+1}^2 \rangle - \langle B_{t+1} \rangle^2 \simeq
{2 N^2 B_t^2}/{\lambda }$,
giving for the diffusion constant of the front position $X_t$ 
\begin{equation}
D \simeq \frac{\langle B_{t+1}^2 \rangle - \langle B_{t+1} \rangle^2}{
\langle B_{t+1} \rangle^2 } \simeq \frac{2 }{\lambda L^2}.
\label{Dnoisy1}
\end{equation}
Compared to the exact result~(\ref{exact}), this has the wrong $L$ dependence
and also depends on the precise value of the cutoff $\lambda$. The
velocity $v_{\rm cutoff} - D/2 $ obtained in this case is also not a
better approximation than the cutoff velocity~(\ref{vcutof1}). We
conclude that the effect of the noise at the left of the cutoff in the
present model is too small to explain the value of the exact diffusion
constant~(\ref{exact}) and can therefore be neglected.

A last approximation is to keep as only source of noise the stochastic
position~$x_{\text{max}}$ of the rightmost
particle\cite{BrunetDerrida.01}: we
choose~$x_\text{max}$ using the distribution of the exact front dynamics
and, on the left of~$x_\text{max}$, we use for~$u_{t+1}(x)$ the average
value of the front \emph{given that the rightmost particle is
on~$x_\text{max}$}. In other words:
\begin{equation}
u_{t+1}(x)=\begin{cases}0&\text{if $x>x_\text{max}$,}\\
		\frac{1}{N}+\frac{N-1}{N}\langle u_{t+1}(x)\rangle
			 &\text{if $x<x_\text{max}$.}
	 \end{cases}
\end{equation}
From (\ref{noisy1-gumbel}), one gets
\begin{equation}
x_\text{max}=\ln\frac{NB_t}{q}\quad\text{with}\quad
\prob(q)=e^{-q}\theta(q),
\end{equation}
and from (\ref{Bt-def}),
\begin{equation}
\frac{B_{t+1}}{B_t}=\frac{N}{q}+(N-1)\int_{q/N}^{+\infty}\frac{1-e^{-u}}{u^2}
\ du.
\end{equation}
Notice that, in that approximation, $\langle B_{t+1}\rangle=\infty$ due
to the contribution of small~$q$ or, equivalently, of
large~$x_\text{max}$. For $q$ small compared to~$N$, one obtains
\begin{equation}
\frac{B_{t+1}}{B_t}=N\left[\frac{1}{q}+\ln N+1-\gamma_E-\ln
q+o\left(1\right)\right].
\label{BB}
\end{equation}
Using the definition~(\ref{DeltaXt-def}) of the displacement~$\Delta X_t$
and the identity~(\ref{integral-representation}), one can compute from
(\ref{BB}) the generating function $\langle \exp(-\delta \Delta X_t)\rangle$.
Up to order $1/\ln N$, one finds nearly the same result as in the exact
solution: one just needs to replace~$1-2\gamma_E$ in (\ref{result}),
by~$2-2\gamma_E$. This means that, up to order~$1/\ln N$, the velocity of
the front in this approximation is shifted by the small amount~$1/\ln N$,
and all the other cumulants are the same as in the exact model.

\medbreak

In this work, we have shown how the problem of directed polymers
with~$N$ sites in $1 + \infty$ dimension can be reduced to a noisy
traveling wave equation (\ref{noisy1}, \ref{noisy2}). For one special
choice of the bond disorder (\ref{gumbel}), we could calculate exactly 
the velocity and diffusion constant (\ref{exact}) of the front, and even
all the cumulants (\ref{result}). The reason which makes this case soluble is
that, at each time step, the only information one needs to keep about the
past is a single variable $B_t$ given by (\ref{Bt-def}). This is similar to
what was observed recently for shocks in exclusion processes
\cite{benAvraham.98, Krebs.03} where, in certain cases, one can decouple
the evolutions of the position of the shock and of its shape. Comparing
several approximation schemes has shown that, in the present case, the
cutoff approximation \cite{BrunetDerrida.97, Kessler.98,
PechenikLevine.99, PanjaVanSaarloos.02} gives a good estimate of the
front velocity, and that the full large $N$ fluctuations of the front can
be obtained by considering the effect of the noise on the rightmost
particle. The predominance of this rightmost particle might be related
to some difficulties noticed in a previously studied growth
model\cite{MarsiliBray.96} where only the first cumulants of the heights
were considered.

The front described by (\ref{noisy1}, \ref{noisy2}) is peculiar
because $N$ appears both in the noise term and the traveling wave
equation itself and because neglecting the noise in~(\ref{noisy1}) leads to
an ill defined traveling wave equation. In the mean field approximation,
one obtains a F-KPP-like front equation~(\ref{thermo}) which still depends
on~$N$ and its velocity diverges like~$\ln N$. These peculiarities make
the problem considered here rather different from usual traveling wave
equations and does not allow us to use the present exact solution to
check the validity of the $\ln^{-2} N$ shift of the velocity and of the
$\ln^{-3} N$ dependence of the diffusion constant which have been
suggested by a number of numerical calculations
\cite{BrunetDerrida.01, Moro.04}. The approximations (cutoff or noise
limited to the rightmost particle) successfully tested here should
however be helpful to describe more standard front equations.

Of course, from both the points of view of the theory of disordered
systems and of the theory of noisy traveling wave equations, it would be
interesting to attack the case of a general distribution~$\rho(s)$. A
starting point
could be to try to make a perturbation theory where the Gumbel
distribution would be a zero-th order approximation. No need to say that
obtaining the~$\ln^{-2}N$ correction to the velocity would require a
delicate resummation of this perturbation theory.

\end{document}